\documentclass[10pt,journal,compsoc]{IEEEtran}
\usepackage{url}
\usepackage{amsfonts}
\usepackage{amsmath}
\usepackage{amsthm}
\usepackage{amssymb}
\usepackage{color}
\usepackage{graphicx} 
\usepackage{epstopdf}
\usepackage{bm}
\usepackage{dcolumn}% Align table columns on decimal point
\usepackage{bm}% bold math
\usepackage{hyperref}% add hypertext capabilities
\usepackage{dsfont}
\usepackage[caption=false]{subfig}
\usepackage{float}

\providecommand{\U}[1]{\protect\rule{.1in}{.1in}}

\graphicspath{{./Figures/}}

\newtheorem{lemma}{Lemma}

\begin{document}

\title{SIS Epidemics in Multilayer-based Temporal Networks }

\author{Aram Vajdi, David Juher, Joan Salda\~{n}a, Caterina Scoglio}

\IEEEtitleabstractindextext{%
\begin{abstract}
To improve the accuracy of network-based SIS models we introduce and study a multilayer representation of a time-dependent network. In particular, we assume that individuals have their long-term (permanent) contacts that are always present, identifying in this way the first network layer. A second network layer also exists, where the same set of nodes can be connected by occasional links, created with a given probability. While links of the first layer are permanent, a link of the second layer is only activated with some probability and under the condition that the two nodes, connected by this link, are simultaneously participating to the temporary link. We develop a model for the SIS epidemic on this time-dependent network, analyze equilibrium and stability of the corresponding mean-field equations, and shed some light on  the role of the temporal layer on the spreading process.
\end{abstract}

}

\maketitle

\IEEEdisplaynontitleabstractindextext

% \fi
\IEEEpeerreviewmaketitle

\section{introduction}
The stochastic susceptible-infected-susceptible (SIS) model over a complex network is a mathematical approach for describing the spread of a pathogen in a population with heterogeneous connectivity among individuals \cite{van2009virus,chakrabarti2008epidemic, pastor2015epidemic, pastor2001epidemic}. Such a stochastic model is suitable when the description of the spreading process at the individual level includes some uncertainty. Indeed, the analysis of the SIS model over static networks has clarified the network structure role in the emergence of the endemic state. This result, in turn, has provided opportunities to control an epidemic by altering the network structure, even though there are multiple sources of uncertainty at the individual level \cite{preciado2013optimal, gusrialdi2018distributed, van2011decreasing}.

In the SIS spreading model, the nodes are either susceptible or infected.
If a node is susceptible, it becomes infected due to interactions with the infected neighbors
in the network, and if it is infected, it can recover and become susceptible again. Although there are articles that consider non-Markovian spreading processes \cite{nowzari2015general, van2013non}, the most
common assumption in the networked spreading literature is that the time duration a node stays
infected is a random variable with an exponential distribution. Moreover, the same assumption is
held for the infection process. In other words, the probability a susceptible node with one infected
neighbor stays susceptible decreases exponentially with time. An important result regarding the SIS model is that the infection in a population dies out exponentially fast if $\beta/\delta<1/\lambda_{max}(A)$ where $\beta,\ \delta$ are the infection transmission and recovery rates, and $\lambda_{max}(A)$ is the largest eigenvalue of the static network adjacency matrix \cite{van2009virus,chakrabarti2008epidemic}. This result is obtained using the N-intertwined equation of the SIS model which approximately describes the SIS model, whose exact mathematical treatment is intractable \cite{van2011n}. In fact, it is shown that the N-intertwined equations provide an upper-bound for the prevalence of infection in the exact SIS process \cite{cator2014nodal,donnelly1993correlation}. Although the aforementioned result is significant for controlling epidemic, the assumption that the underlying network is
known and static is not justifiable in some important instances of real-world populations. For example, the contacts resulting from the current trend in online dating cannot be represented as a static network. This motivates our work to analyze the SIS spreading process over time-varying networks.

In fact, in the existing literature, we can find several works analyzing  the SIS processes over various models of dynamic networks \cite{zhang2017random,holme2012temporal,georgiou2015solvable}. Par$\acute{\text{e}}$ \textit{et al.} analyze the N-intertwined approximation of the SIS process when the adjacency matrix of the network is a deterministic and continuous function of time \cite{pare2018epidemic}. Another approach to model temporariness of contacts is to adopt the switching network concept.  In such a model the contact network randomly switches among a set of predetermined adjacency matrices. In \cite{rami2014stability, sanatkar2016epidemic} the authors have studied sufficient conditions for stability of the disease free equilibrium in the SIS spreading model over switching networks. Another class of time-varying network that has been studied in the existing literature is the edge-Markovian networks where the edges appear and disappear following independent Markov processes \cite{clementi2010flooding}. In \cite{taylor2012epidemic}, the authors have used an improved effective degree compartmental
modeling framework to study the SIS spreading process in the edge-Markovian networks. Ogura et al. consider a generalized version of edge-Markovian model where the inter-event time distribution for the appearance and disappearance of the links is not necessarily exponential \cite{ogura2016stability}. Moreover, they provide a sufficient condition for exponential stability of the disease-free state in the SIS process that is unfolding on such a time-varying network. A different approach to model time-varying networks is the activity driven network, which has been studied mostly in the physics literature \cite{perra2012activity,pozzana2017epidemic}. Typically, in a discrete time activity driven model, nodes  become active and establish links at each time step with some randomly selected nodes in the population. Moreover, active nodes cut their existing links randomly with some probability. However, in some practical cases such a model is an oversimplification of the real scenario, and may miss some critical aspects for the infection spreading in a population. 

In this paper, we study the SIS spreading process over time-varying networks. First, we propose a new framework for the modeling of
temporal networks, using the concept of layer of potential contacts. We assume a potential
link becomes an active contact with a link-specific probability, if the nodes on both ends of the potential link are active. In this approach, the temporal contacts result from the transition of nodes between active and inactive states. Second, we develop a mean-field type approximation to describe the SIS spreading process over such a temporal network. Moreover, we discuss why such approximation is relevant to the exact description of the process. Third, we analyze the disease-free state of the SIS spreading process using the mean-field equations and we find a condition that guarantees the exponential die out of an infection in a time-varying network that can be described via our modeling approach. Finally, using the the exact simulation of the process, we show how the duration of potential links can affect the metastable state of the SIS spreading process.

Our motivation in this paper is to provide a theoretical background for controlling the propagation of sexually transmitted diseases, which are well represented by the SIS model \cite{juher2017network}. For such processes, we need to consider the fact that an individual may have a permanent partner and occasional partners as well. Moreover, occasional partners are not found randomly among the whole population, but only within a subset of individuals sharing some kind of affinity. Hence, in our model of time-varying network, we account for a permanent contact layer beside the potential layer. Moreover, the potential layer is quantified by the probabilities that the nodes may develop a link. Our analysis shows that we can still contain the infection spreading by satisfying a condition that guarantees the exponential die out of infection, even though there are numerous uncertainty in such a system. Such a sufficient condition depends on different parameters that describe our model.  

\section{network and spreading models} 
In the following, we first introduce the notation and the assumption of the two-layer temporal network model, and later, we develop the SIS mean field-equations on this network model.

\subsection{Two-layer temporal network model}\label{netmodel}
\begin{figure}[t]
\centering
\includegraphics[width=1 \columnwidth]{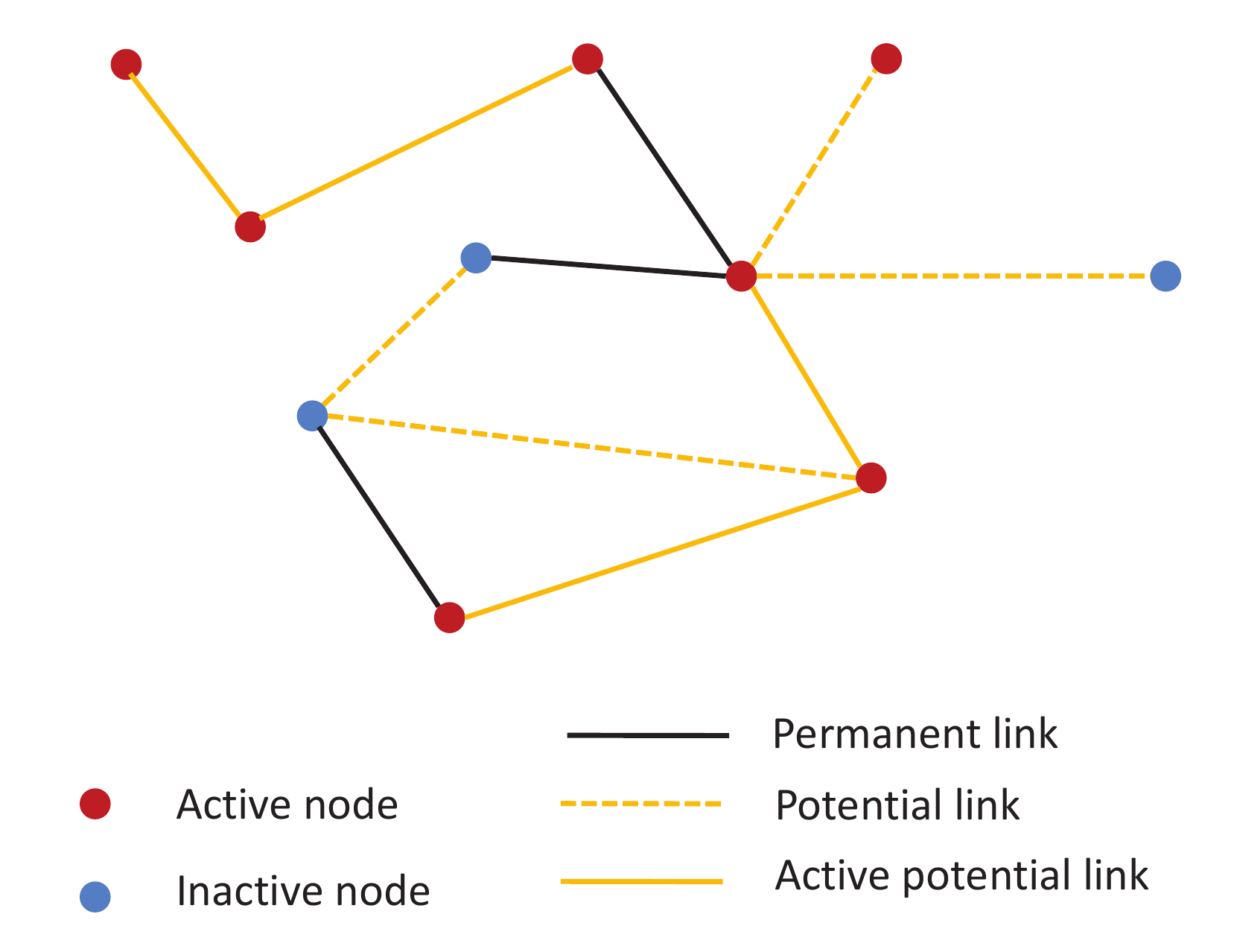}
\caption{A snapshot from a realization of the network model. At any time $t$ the nodes are either active or inactive. A potential link is activated with probability $p_{0}$ if both ends of it are active at the same time.}
\label{netmodelfig}
\end{figure}

We consider a population of $N$ agents that are connected with two different types of links. The first network layer, $\mathbb{L}_{1}$, represents permanent contacts (long term relationships) among the agents. Beside these permanent links, we assume a second type of links that are potential contacts and they become active contacts with a probability $p_{0}$ only when the agents on both sides of the links are simultaneously seeking occasional partners. This second layer of links is denoted by $\mathbb{L}_{2}$. In general, $p_{0}$ can be different for each pair of nodes. However, since it is straightforward to generalize our result to the heterogeneous case, we assume the same  $p_{0}$ value for all potential links. By definition, the intersection of the two network layers is empty. While the links in layer $\mathbb{L}_{1}$ always can transmit infection, a link in layer $\mathbb{L}_{2}$ transmits the infection only when it becomes an active contact. In our model the activation of a potential link in $\mathbb{L}_{2}$ depends on the activity state of agents at both ends of the link. Apart from the node infection state, we assume  $t$ the individuals are either active or inactive at any time. When a node becomes active, it seeks contact among the active neighbors in $\mathbb{L}_{2}$ and with a probability $p_{0}$ it activates an occasional contact. Later, when one of the two nodes goes to the inactive state, the occasional contact is inactivated. This node transition between active and inactive states introduces temporariness in the contact network. Here, we assume node activation processes are independent Poisson processes, where node $i$ becomes active with rate $\gamma_{1}^{i}$, and if it is active, it goes to the inactive state with rate $\gamma_{2}^{i}$. Since the inverse of the transition rate is the expected value of transition time, if node $i$ is active, it is expected to stay active for a period of time of length $(\gamma^{i}_{2})^{-1}$. Thus, when we want to model a node that is frequently activating occasional links, we can assign high values of $\gamma_{2}$ and $\gamma_{1}$ to that node. Moreover, if a node does not participate in the occasional contacts ---it never becomes active--- $\gamma_{1}$ is set equal to zero for that node. Figure \ref{netmodelfig} shows a snapshot of a realization of the temporal network.

Since in this model the inactivation time for each node has an exponential distribution, and the inactivation of a temporal contact depends on the both ends of it, it is straightforward to see the temporal contact duration has an exponential distribution. In fact, a temporal contact disappears the moment one end of the link becomes inactive. Since the minimum of two independent random variables with exponential distributions is distributed exponentially with a rate that is summation of the rates in the independent distributions, we can deduce the duration of a temporal contact between nodes $i$ and $j$ has an exponential distribution with the rate $\gamma_{2}^{i}+\gamma_{2}^{j}$. Hence, the expected duration of the contact is $(\gamma_{2}^{i}+\gamma_{2}^{j})^{-1}$.    
\subsection{SIS epidemics on two-layer temporal networks}\label{comp}
In this section we develop a mean-field type approximation to describe the spreading of infection on the temporal network introduced in section \ref{netmodel}. Next, we discuss the relevance of such approximation to the exact spreading process.  

The susceptible-infected-susceptible (SIS) model is a popular approach for studying infection spreading. In this model, each node is either susceptible (S) or infectious (I). We assume the infection and recovery processes are Poisson processes, where an infectious node recovers with a rate $\delta$ and propagates the contagion to a susceptible neighbor with a rate $\beta$. When a susceptible node is in contact with several infectious nodes, it is assumed each infected neighbor acts independently. Thus, the susceptible node contracts the infection with a rate that is the sum of the rates of all the independent infection processes. 

Combining the network model and the SIS spreading process, we deduce each node can assume one of four different states: $\mathcal{S}_{1}$ susceptible and inactive, $\mathcal{S}_{2}$ susceptible and active, $\mathcal{I}_{1}$ infectious and inactive, $\mathcal{I}_{2}$ infectious and active. If  $S^{i}_{1}$, $S^{i}_{2}$, $I^{i}_{1}$ and $I^{i}_{2}$ represent the probabilities that the node $i$ is in one of the four states in the mean-field approximation, the equations for the time evolution of $S^{i}_{1}$, $S^{i}_{2}$, $I^{i}_{1}$ and $I^{i}_{2}$ can be written as
%\begin{equation}\label{eqa}
%\begin{split}
\begin{subequations}\label{eqa}
\begin{align}
{\dot{S}_{1}}^{i}&=-\gamma_{1}^{i} S_{1}^{i}+\gamma_{2}^{i}S_{2}^{i}+\delta I_{1}^{i}-\beta\sum_{j}a_{1}^{ij}S_{1}^{i}(I_{1}^{j}+I_{2}^{j}),\label{eqaa}
\\
{\dot{I}_{1}}^{i}&=-\gamma_{1}^{i} I_{1}^{i}+\gamma_{2}^{i}I_{2}^{i}-\delta I_{1}^{i}+\beta\sum_{j}a_{1}^{ij}S_{1}^{i}(I_{1}^{j}+I_{2}^{j}),\label{eqac}
\\
{\dot{S}_{2}}^{i}&=-\gamma_{2}^{i} S_{2}^{i}+\gamma_{1}^{i}S_{1}^{i}+\delta I_{2}^{i}-\beta\sum_{j}a_{1}^{ij}S_{2}^{i}(I_{1}^{j}+I_{2}^{j})\label{eqab}\\&-\beta^{\prime}\sum_{j}a_{2}^{ij}S_{2}^{i}I_{2}^{j},\nonumber
\\
{\dot{I}_{2}}^{i}&=-\gamma_{2}^{i} I_{2}^{i}+\gamma_{1}^{i}I_{1}^{i}-\delta I_{2}^{i}+\beta\sum_{j}a_{1}^{ij}S_{2}^{i}(I_{1}^{j}+I_{2}^{j})\label{eqad}\\&+\beta^{\prime}\sum_{j}a_{2}^{ij}S_{2}^{i}I_{2}^{j},\nonumber
\end{align}
\end{subequations}
%\end{split}
%\end{equation}
where $\beta^{\prime}=p_{0}\beta$.
In the equations above, $a^{ij}_{1}$ is an element of the adjacency matrix $A_{1}$ for the permanent contact layer $\mathbb{L}_{1}$ with $a^{ij}_{1}=1$, if the nodes $i$ and $j$ are in permanent contact, and $a^{ij}_{1}=0$ otherwise. Similarly, $a^{ij}_{2}$ is the $(i,j)$ element of the adjacency matrix $A_{2}$ corresponding to the potential-contact layer $\mathbb{L}_{2}$. \textbf{It is important to note} that, when $p_{0}$ has different values for each pair, we can absorb $p_{0}$ in the adjacency matrix $A_{2}$ and the element of the $A_{2}$ matrix become the pair-specific probabilities of developing contacts.

Equation (\ref{eqaa}) describes how the probability of node $i$ being in the state $\mathcal{S}_{1}$ changes with time. The first term on the r.h.s. of the equation reflects the fact that the inactive susceptible node $i$ becomes active with a rate $\gamma^{i}_{1}$ and the second term indicates if the node $i$ is in the state $\mathcal{S}_{2}$ it goes to the inactive state with the rate $\gamma^{i}_{2}$. The third term originates from the recovering process of inactive infected nodes. In the forth term, each addend is the multiplication of the probability that the node $i$ is inactive susceptible and the probability that a permanent neighbor of node $i$ is infected. 

In equation (\ref{eqab}), we take into account the two different sets of neighbors that propagate infection to the active susceptible node $i$. The forth term on the r.h.s. of this equation arises from the contagion propagation by the infectious permanent neighbors of node $i$. In the fifth term, the summation is over the multiplication of the probability that the node $i$ is in the state $\mathcal{S}_{2}$ and the probability that a potential neighbor of node $i$ in the activity layer $\mathbb{L}_{2}$, is infectious and also active. When the nodes $i$ and $j$ are active and they are neighbors in the activity layer $\mathbb{L}_{2}$, they develop a link with probability $p_{0}$. Hence, the summation in the fifth term of this equation is multiplied by $p_{0}$.

\begin{figure}[t]
\centering	
	\subfloat[]{
		\includegraphics[width=.455\columnwidth]{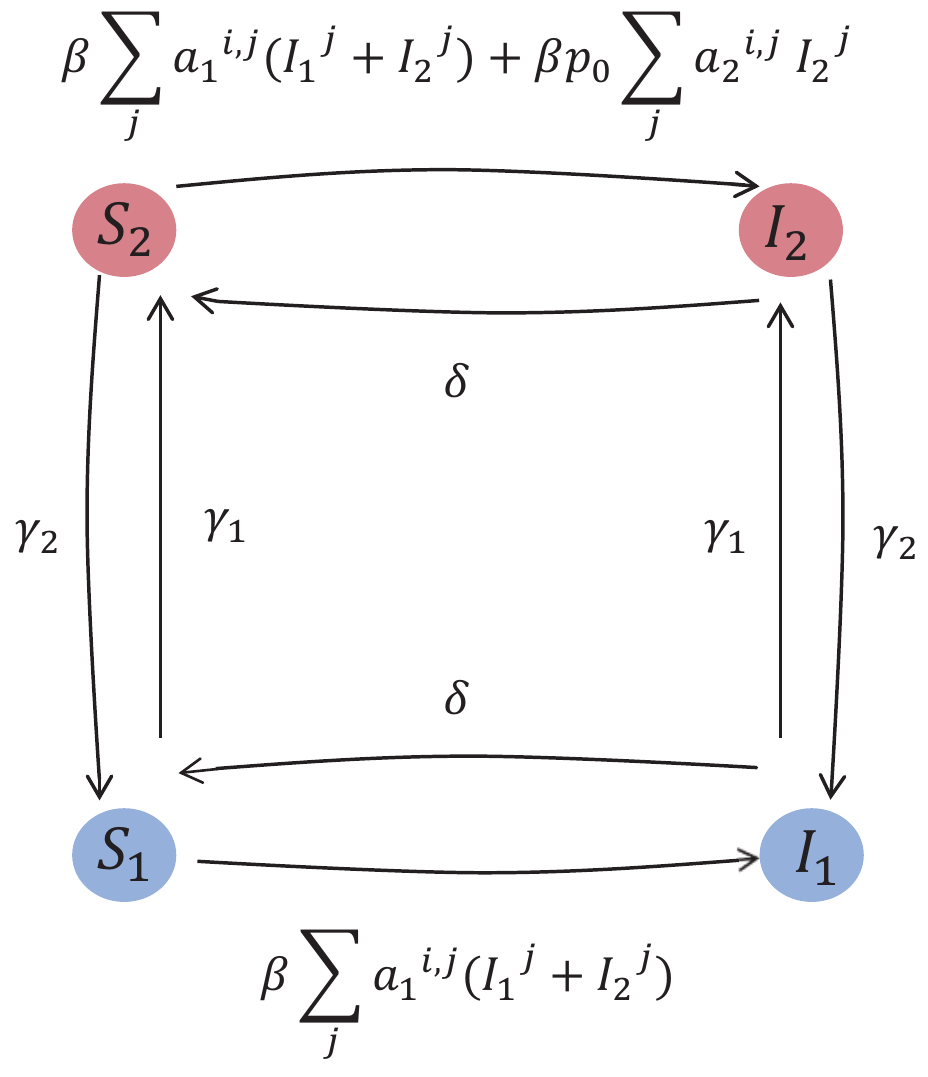}
\label{markov}} \ 
	\subfloat[]{ 
	
		\includegraphics[width=.47\columnwidth]{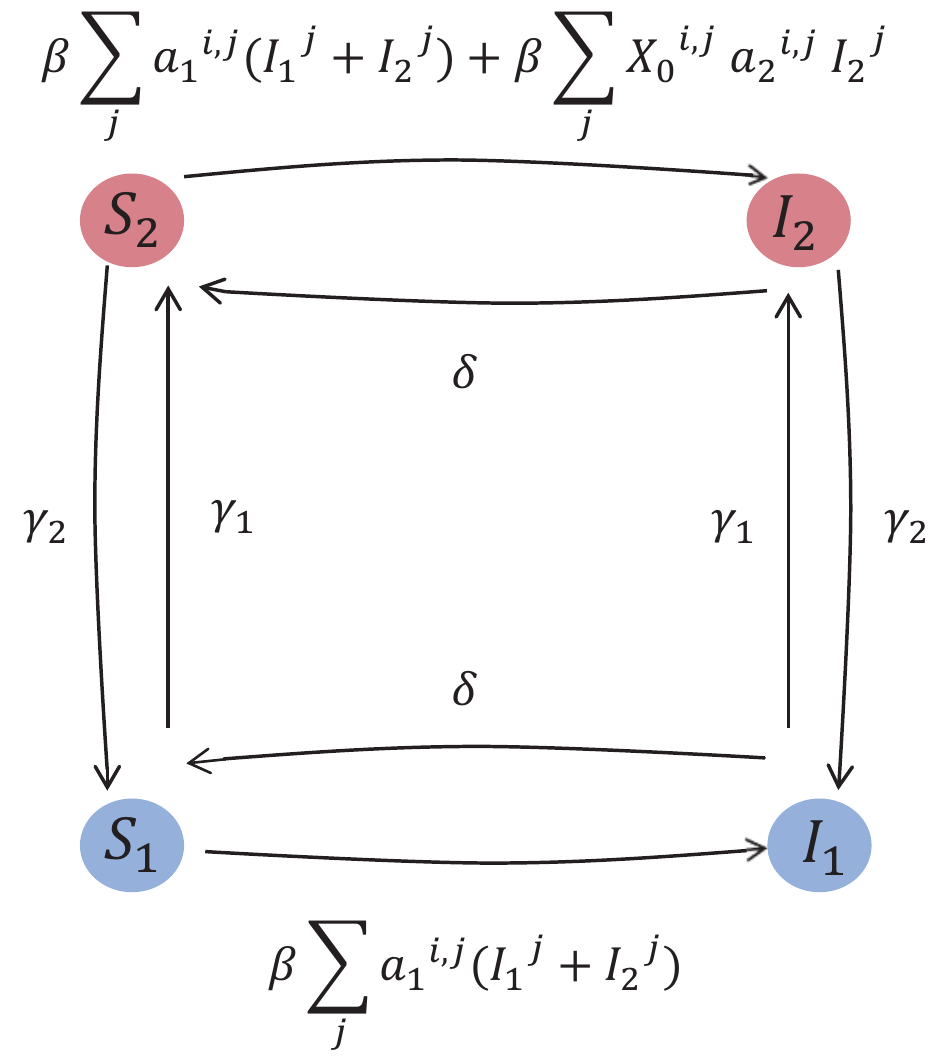}
\label{exact}}
		\caption{The figures show the diagrams of node transitions among different node states. The rates of each transition is specified on the arrow that indicates the transition. (a) shows diagram of the Markov process which is discussed in section \ref{comp}, and (b) shows diagram of the exact process. In these figures $I_{1}^{j}=1$ ($I_{2}^{j}=1$) if node $j$ is infected and inactive (active), otherwise it is zero. In diagram (b) $X_{0}^{i,j}$ is a Bernoulli random variable that has value one with probability $p_{0}$. This random variable is drawn each time a pair of active nodes $(i,j)$ with a potential link between them occurs, regardless of their disease status.} 
	\label{model}%
\end{figure}

Equations \ref{eqa} describe approximately the exact (stochastic) spreading model. Our numerical simulations show these equations lead to nodal infection probabilities that are upper bounds for the infection probabilities in the exact spreading model. We conjecture this is a general property of this model. In the following, we give an intuitive picture to justify our conjecture that equations \ref{eqa} provide an upper bound for the exact process. Readers familiar with continuous-time Markov chain and the mean-field approximation of SIS process over static one-layer network \cite{sahneh2013generalized,van2009virus} may recognize that the equations \ref{eqa} are the N-intertwined approximation of a continuous Markov processes similar to our model but with a difference. In contrast to the exact description of our model, for this Markov process a link in layer $\mathbb{L}_{2}$ is activated whenever the nodes at both ends of the link are active with the infection transmission through the link being $\beta^{\prime}=p_{0}\beta$ instead of $\beta$. Figure \ref{markov} shows the nodal transitions in the Markov process. However, in our model, when both ends of a link are active, the link becomes activated with probability $p_{0}$, and transmits infection with rate $\beta$ if one of the nodes is infected. Figure \ref{exact} shows the nodal transitions in our model. Our simulations show that the equations \ref{eqa} give an upper bound for the nodal infection probabilities in the Markov process described above, and that the nodal infection probabilities in this Markov process are higher than that of our stochastic model. 

To justify our conjecture that the equations \ref{eqa} gives an upper bound for the nodal infection of the above mentioned Markov process we invoke the intuitive argument in \cite{cator2014nodal}, where the authors prove the Markovian SIS process over a static one-layer network is upper-bounded by the N-intertwined approximation. In fact, equation \ref{eqac} would be an exact equation for the Markov process if we replace in this equation $S_{1}^{i}(I_{1}^{j}+I_{2}^{j})$  with $\Pr(x_{i}=\mathcal S_{1}, x_{j}=\mathcal I_{1}\ \text{or} \ \mathcal I_{2})$, which is the joint probability that node $i$ is inactive and susceptible, and node $j$ is infected. Moreover, since two neighboring nodes can only enhance the infection probabilities of each other and their activity states are independent, we expect the infection states would be non-negatively correlated. In other words, when we know node $j$ is infected the expectation to observe node $i$ in the susceptible state is less than the case when we do not know the state of node $j$, 
\[
\Pr(x_{i}=\mathcal S_{1} | x_{j}=\mathcal I_{1}\ \text{or} \ \mathcal I_{2})\leq  \Pr(x_{i}=\mathcal S_{1}).
\] 
If we rewrite the inequality above as
\[
\Pr(x_{i}=\mathcal S_{1} , x_{j}=\mathcal I_{1}\ \text{or} \ \mathcal I_{2})\leq  S_{1}^{i}(I_{1}^{j}+I_{2}^{j}),
\]
we can see the summands in equation \ref{eqac} are upper-bounds for the corresponding terms, $\Pr(x_{i}=\mathcal S_{1}, x_{j}=\mathcal I_{1}\ \text{or} \ \mathcal I_{2})$, in the exact equation for the Markov process. Since these terms appears with positive sign, they only increase the infection probability. Using a same argument about the correlation of nodal infection in equation \ref{eqad}, we expect the N-intertwined approximation in equation \ref{eqa} gives an upper bound for the nodal infection probabilities in the Markov process and our simulations show that it is in fact an upper bound. In order to compare the nodal infection probabilities in the Markov model and the exact description of our stochastic model, consider an instance where at $t_{1}$ one end of an $\mathbb{L}_{2}$ link is active susceptible while the other end is active infected. If $t_{2}$ is the later instant when either the infectious node recovers or one of the nodes becomes inactive, in the Markov process, the probability for transmission of infection through the link is $1-e^{-p_{0} \beta (t_{2}-t_{1})}$. But in our model this probability of transmission is $p_{0}(1-e^{-\beta (t_{2}-t_{1})})$ which is always smaller than that of the Markov model. Thus, we expect the infection probabilities in our model will be upper-bounded by the probabilities from the Markov process which are in turn smaller than the values obtained  from the N-intertwined approximation in equation \ref{eqa}. This property of equations \ref{eqa} is particularly useful in controlling the infection spreading. In fact, if any initial infection that is governed by equation \ref{eqa} dies out we know that the infection can not survive in our model. 

\section{The  mean-field model on regular random networks}

To study analytically the impact of the transition rates $\gamma^i_1$ and $\gamma^i_2$ between layers on the epidemic spread, we consider the case where $\mathbb{L}_{1}$ and $\mathbb{L}_{2}$ are regular random networks of degree $k_1$ and $k_2$, respectively. Moreover, let us assume that all the nodes have the same transition rates, i.e. $\gamma^{i}_j = \gamma_j \, \forall \, i$ ($j=1,2$). This means that, for any node, the probability of being in layer $2$ is $p_2 = \gamma_1/(\gamma_1+\gamma_2)$, and similarly for layer 1 ($p_1 = \gamma_2/(\gamma_1+\gamma_2)$). Hence, $S^i_j = p_j - I^i_j$ ($j=1,2$). Introducing this relation in the previous system and summing the equations for the infected nodes in each layer, we have  
\begin{eqnarray*}
{\dot{I}_{1}}  & = & (\beta k_1 p_1 - (\gamma_1+\delta)) I_1 + (\beta k_1 p_1 + \gamma_2)I_2 
\\ & & - \beta k_1 \sum\limits_{j} \left( \sum\limits_{i} a_{1}^{ij} I_{1}^{i} \right) (I_{1}^{j} + I_{2}^{j})
\\ & \\
{\dot{I}_{2}}  & = & (\beta k_1 p_2 + \gamma_1) I_1 + \left( \beta p_2 ( k_1 + p_0 k_2) - (\gamma_2 + \delta) \right) I_2 
\\ & \\ 
& \hspace*{-1cm} - & 
\hspace*{-0.75cm} 
\beta \sum\limits_{j} \left( \sum\limits_{i} a_{1}^{ij} I_2^i \right) (I_{1}^{j} + I_{2}^{j})
- \beta p_0 \sum\limits_{j} \left( \sum\limits_{i} a_{2}^{ij} I_{2}^{i} \right) I_{2}^{j}, 
\end{eqnarray*}
where $I_{1} = \sum_i I_{1}^{i}$ and $I_{2} = \sum_i I_{2}^{i}$ are the expected number of infected nodes in layer 1 and layer 2, respectively. Let us now approximate the sums $\sum\limits_{i} a_{l}^{ij} I_{l}^{i}$ by $k_l I_l/N$, which is a good approximation as long as the degree distribution has low variance (as in regular random networks or \"Erdos-R\'eny networks) and the mean degree is high. Then, after dividing both sides of the equations by $N$, we have the following system of equations for the disease prevalence $\rho_j=I_j/N$ in each layer:
\begin{eqnarray}
\dot{\rho}_1 & = & (\beta k_1 p_1 - (\gamma_1+\delta)) \rho_1 + (\beta k_1 p_1 + \gamma_2) \rho_2  \nonumber \\
& & - \beta k_1 \rho_1 (\rho_1+\rho_2)
\\ & \nonumber \\
\dot{\rho}_2 & = & (\beta k_1 p_2 + \gamma_1) \rho_1 + (\beta p_2(k_1+p_0 k_2) - (\gamma_2+\delta)) \rho_2  \nonumber \\
& & - \beta \rho_2( k_1 (\rho_1 + \rho_2) + p_0 k_2 \rho_2).
\end{eqnarray}

To study the linear stability of the disease-free equilibrium (DFE), we consider the Jacobian matrix of the previous system around the DFE
$$
J_0 = \begin{pmatrix}
\beta k_1 p_1 - (\gamma_1+\delta) & \beta k_1 p_1 + \gamma_2 \\
\beta k_1 p_2 + \gamma_1 & \beta p_2(k_1+p_0 k_2) - (\gamma_2+\delta)
\end{pmatrix}.
$$
One can see that the discriminant $\Delta$ of the characteristic equation $\det(J_0 - \lambda I)=0$ is always positive. Precisely, after some algebra and using that $p_1+p_2=1$, we end up with  
$$
\Delta = (\beta(k_1 - k_2 p_0 p_2) + \gamma_1+\gamma_2)^2 + 4 \beta k_2 p_0 p_2 (\gamma_1 + \beta k_1 p_2) > 0,
$$
which implies that $J_0$ has two distinct real eigenvalues $\lambda_1 > \lambda_2$. Therefore, to guarantee that $\lambda_1$ traverses $0$ when using a tuning parameter of interest, we need that ${\rm trace}(J_0) = \beta k_1 - (\gamma_1+\delta) + \beta k_2 p_0 p_2 - (\gamma_2 + \delta)  < 0$. This condition implies that, at least in one layer, the corresponding basic reproduction number $R^{(j)}_0 < 1$ with $R^{(1)}_0=\beta k_1/(\gamma_1+\delta)$ and $R^{(2)}_0= \beta k_2 p_0 p_2 / (\gamma_2 + \delta)$. Then, the condition for $\lambda_1=0$ follows from $\det(J_0)=0$ which is equivalent to 
$$
\beta k_2 p_0 p_2 (\beta k_1 p_1 - (\gamma_1+\delta)) = (\beta k_1 - \delta)(\gamma_1+\gamma_2+\delta),
$$
which requires $\gamma_1,\gamma_2 > 0$ if $\beta k_1 \ne \delta$.

\begin{figure}[t] 
\centering
		\includegraphics[width=0.7 \columnwidth]{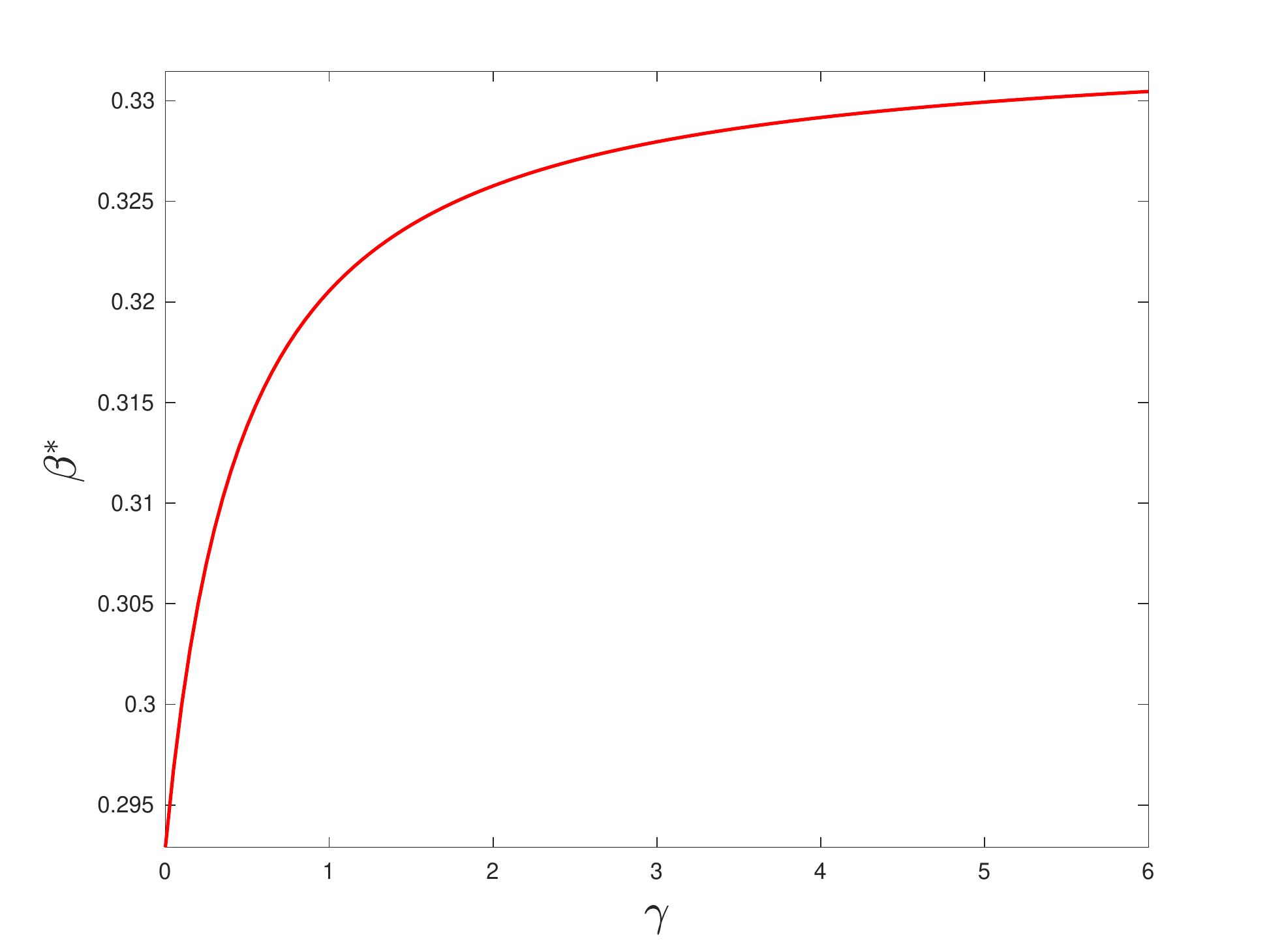}
\caption{Critical value of $\beta$ as a function of $\gamma_1$ in regular random networks. Parameters: $k_1=4$, $k_2=50$, $p_0=0.5$, $\delta=1$, $\gamma_2=\gamma_1$. } 
\label{critbeta}
\end{figure}

The previous condition defines a second degree equation for the critical value of $\beta$, $\beta^*$. It is easy to see that this equation has two real roots $0 < \beta_1 < \beta_2$. Since we want the value of $\beta$ for which $\lambda_1$ goes from negative to positive, $\beta^*=\beta_1$.  Fig.~\ref{critbeta} shows the dependence of $\beta^*$ with the transition rate $\gamma_1$ obtained by solving the previous equation for $\gamma_1=\gamma_2$. So, in this figure, the probability for a node of being in $\mathbb{L}_{2}$ is always 1/2. However, although a node always spends half of its time with contacts in $\mathbb{L}_{2}$, how it visits this layer (short and frequent visits or longer but less frequent ones) affects the spread of the disease. 

\begin{figure}[t] 
\centering
		\includegraphics[width=0.7\columnwidth]{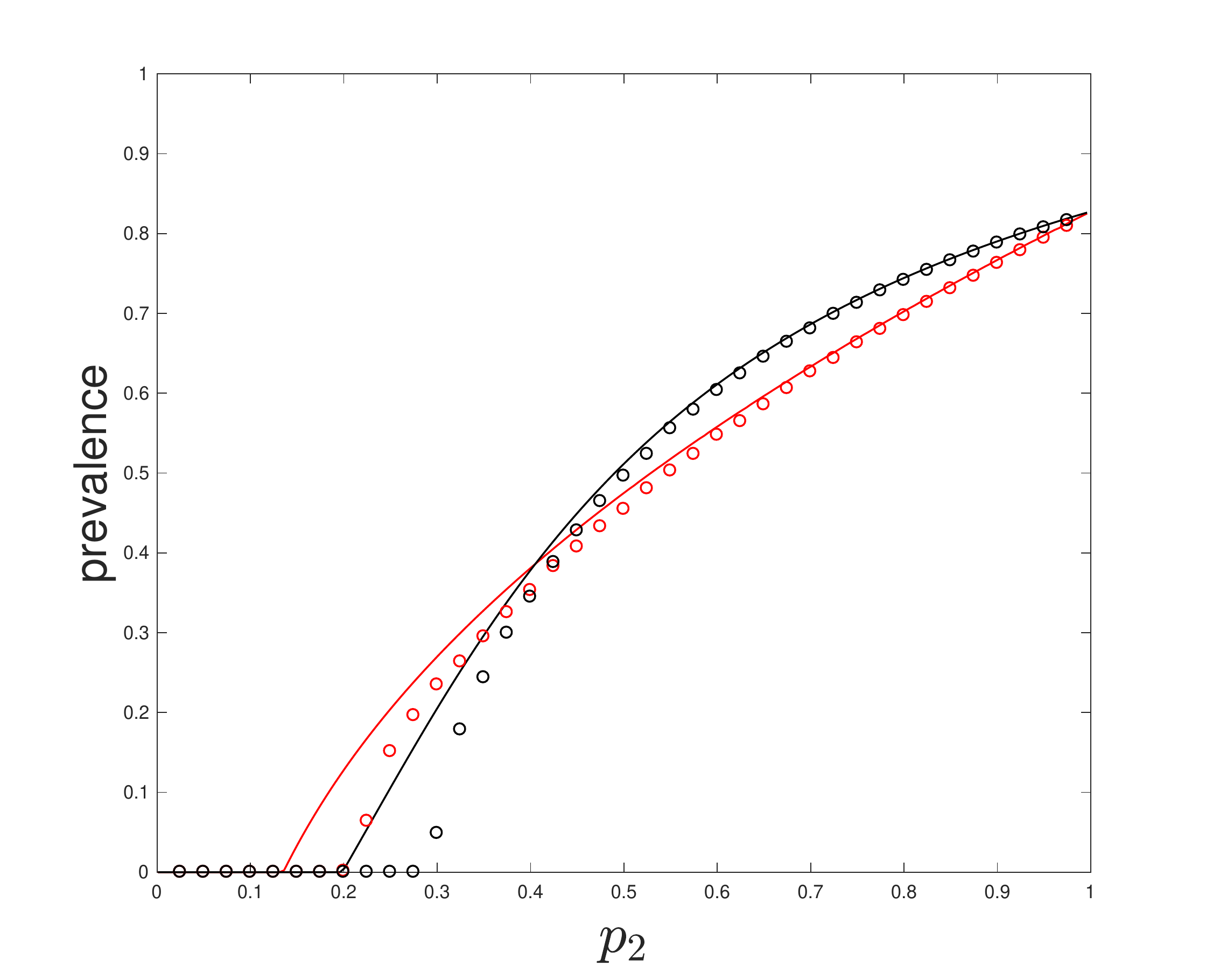}
\caption{Disease prevalence as a function of $p_2$ in regular random networks. Circles show, for each set of parameters values, the mean prevalence in networks of size 500 after 1000 runs of the Markov process approximated by the mean-field model. Parameters: $k_1=4$, $k_2=50$, $p_0=0.5$, $\beta=0.2$, $\delta=1$, $\gamma_1=0.01$ (black), $\gamma_1=10$ (red).} 
\label{crossing}
\end{figure}

A second feature of the MF model is the possibility of having a lower prevalence at the endemic equilibrium for values of $\gamma_1$ leading to lower epidemic thresholds. We illustrate that in Fig.~\ref{crossing} where a bifurcation curve from the DFE is shown using the probability of being in $\mathbb{L}_{2}$, $p_2$, as a tuning parameter. As expected, the higher this probability is, the higher the prevalence because more transmission routes for the infection are used. However, the figure also shows a more surprising fact: $\gamma_1=0.01$ leads to a lower $p_2$ threshold value when compared to $\gamma_1=10$ but, at the same time, it also leads to a lower equilibrium prevalence for $p_2 > 0.37$. We can also observe this feature of the solutions in the output of the simulations over regular random networks of the Markov process corresponding to the mean-field model (see section \ref{comp}). These simulations have been done using Gillespie algorithm until a final time T=600. Finally, figure~\ref{crossing} also reveals that the MF model underestimates the epidemic threshold observed from the stochastic simulations. As discussed in section \ref{comp}, this is due to the higher infection probabilities assumed under the mean-field approach.  

\section{the disease-free equilibrium}\label{treshold} 
In this section we focus on the stability analysis of the disease-free equilibrium of the dynamical system in equation \ref{eqa}, and we find a condition that guarantees the exponential die out of any small initial infection that is introduced in the population. In fact, a bifurcation analysis similar to the one in \cite{van2009virus,darabi2014spreading} shows that, when this condition is not satisfied, there exists another equilibrium state that it is not disease-free. 

For the dynamical system  \ref{eqa}, it is a straightforward observation that the disease-free state given by
\begin{equation}\label{eqb}
S_{1}^{i}=p_{1}^{i}=\frac{\gamma_{2}^{i}}{\gamma_{2}^{i}+\gamma_{1}^{i}},\ \ 
S_{2}^{i}=p_{2}^{i}=\frac{\gamma_{1}^{i}}{\gamma_{2}^{i}+\gamma_{1}^{i}},\ \ I_{1}^{i}=0,\ \ I_{2}^{i}=0,
\end{equation}
is an equilibrium state. In equation \ref{eqb}, $p_{1}^{i}$ and $p_{2}^{i}$ are the probabilities that node $i$ is active and inactive, respectively, at the steady-state of the continuous-time Markov chain that governs the activity of node $i$. Here, we study the evolution of the initial infection around the disease-free equilibrium using the corresponding linearized version of dynamical system \ref{eqa}. In the analysis that comes later, we use set of state variables $I^{i}=I^{i}_{1}+I^{i}_{2}$ and $I^{i}_{2}$ instead of $I^{i}_{1}$, $I^{i}_{2}$. Particularly, this choice of variables directly leads to a relation between the network structure and the model parameters such that, if it is satisfied, the disease-free equilibrium is exponentially stable. If we choose $I^{i}_{1}$, $I^{i}_{2}$, we would need extra algebraic manipulation to get the same relation. 

If $\mathbf{I}^{i}$ and $\mathbf{I}_{2}^{i}$ represent small perturbations from the disease-free equilibrium, using the linearized version of equation \ref{eqa} we obtain the following linear dynamical system 
\begin{subequations}\label{eqd}
\begin{align}
{\dot{\mathbf{I}}}^{i}&=-\delta \mathbf{I}^{i}+\beta\sum_{j}a_{1}^{ij}\mathbf{I}^{j}+\beta^{\prime}\sum_{j}a_{2}^{ij}p_{2}^{i}\mathbf{I}_{2}^{j},
\\
{\dot{\mathbf{I}}_{2}}^{i}&=-(\gamma_{2}^{i}+\gamma_{1}^{i}) \mathbf{I}_{2}^{i}+\gamma_{1}^{i}\mathbf{I}^{i}-\delta \mathbf{I}_{2}^{i}+\beta\sum_{j}a_{1}^{ij}p_{2}^{i}\mathbf{I}^{j}\\&+\beta^{\prime}\sum_{j}a_{2}^{ij}p_{2}^{i}\mathbf{I}_{2}^{j},\nonumber
\end{align}
\end{subequations}
that determines the evolution of the state variables 
\[
X=(\mathbf{I}^{1}, \cdots, \mathbf{I}^{N},\mathbf{I}^{1}_{2}, \cdots, \mathbf{I}^{N}_{2}).
\] 
We can write equations \ref{eqd} as $\dot{X}=JX$ where $J=B-D$ with
\[
B = 
 \begin{pmatrix}
 \beta A_{1} & \beta^{\prime} p_{2} A_{2}  \\
  \beta p_{2} A_{1}+\gamma_{1} & \beta^{\prime}p_{2} A_{2}  
 \end{pmatrix},\  \  D=
\begin{pmatrix}
  \overline{\delta} & 0  \\
  0 & \overline{\delta}+\gamma_{1}+\gamma_{2} 
 \end{pmatrix}.
\]
In the definition of matrices $B$ and $D$ above, $p_{2}, \gamma_{1},\gamma_{2},\overline{\delta}$, are diagonal matrices and the diagonal entries of the latter are the corresponding parameters for different nodes. It is well known that the linear system is stable if $\alpha(J)<0$, where
\[
\alpha(J)=\max\{\Re(\lambda)| \lambda\in \text{spectrum of}\  J\}.
\]

In the following we show there is a threshold  $\beta^{\ast}$ such that for any value of transmission rate  $\beta<\beta^{\ast}$ the disease-free equilibrium is exponentially stable, i.e. $\alpha(J)<0$.

\begin{lemma}\label{lem1}
If the nonnegative matrix $B$ is irreducible, \\
a) there is a real eigenvalue of $J$, denoted by $\lambda_{\max}(J)$, such that any other eigenvalue $\lambda$ satisfies $\Re(\lambda)\leq\lambda_{\max}(J)$, and the eigenvector $Z$ corresponding to $\lambda_{\max}(J)$ is unique and positive, $Z>0$.\\
b)  $\min_{i}\sum_{k}J_{ik}\leq\lambda_{\max}(J)\leq \max_{i}\sum_{k}J_{ik}$ \\
c) If there exists a vector $X\geq0$ such that 
$JX\leq\mu X$, then $X>0$  and $\lambda_{\max}(J)\leq\mu $ with 
$\lambda_{\max}(J)=\mu $ if and only if $X$ is a multiple of $Z$.
\end{lemma}

\proof
From the definition of $J$, we have $J=B-D$ where $B$ is a non-negative matrix and $D$ is a nonnegative diagonal matrix. If we assume $\tau=\max_{k} D_{kk}$ then matrix $C=B-D+\tau I$, with $I$ denoting the identity matrix, is also nonnegative. Under the condition that $B$ 
is irreducible $C$ becomes irreducible. Now we can use Perron-Frobenius theorem for non-negative irreducible matrices \cite{sternberg2010dynamical} to show the statements of lemma \ref{lem1} hold for the matrix $C=J+\tau I$. Since the eigenvectors of $J$ are similar to the eigenvectors of $C$ and the eigenvalues of $J$ can be obtained by subtracting $\tau$ from the eigenvalues of $C$, we deduce the statements of lemma \ref{lem1} also hold for $J$.
\endproof

If we assume $\beta^{\ast}$ is the transmission rate for which $\lambda_{\max}(J_{\beta^{\ast}})=0$ and $Z_{\beta^{\ast}}>0$ is the corresponding eigenvector, using lemma \ref{lem1} it is straightforward to show for any $\beta<\beta^{\ast}$ we have $J_{\beta}Z_{\beta^{\ast}} \leq 0$. Next, we can use the last part of lemma \ref{lem1} and conclude $\lambda_{\max}(J_{\beta})< 0$. This shows that, if $\beta<\beta^{\ast}$, the disease-free equilibrium is exponentially stable. Moreover, to prove the existence of $\beta^{\ast}$, we can use statement ($b$) of lemma \ref{lem1} and consider the limiting cases $\beta\rightarrow 0$ and $\beta\rightarrow \infty$ to show that there are $\beta_{1}$ and $\beta_{2}$ such that $\lambda_{\max}(J_{\beta_{1}})< 0$ and $\lambda_{\max}(J_{\beta_{2}})>0$. Since $\lambda_{\max}(J_{\beta})$ is a continuous function of $\beta$ there should be a 
$\beta^{\ast}$ such that $\lambda_{\max}(J_{\beta^{\ast}})=0$.

In lemma \ref{lem1} we have assumed that the nonnegative matrix $B$ is irreducible. In general, irreducibility of nonnegative matrices can be interpreted as a connectivity condition upon a certain associated graph. For matrix $B$ we can associate a directed graph, $G_{B}$, with $2N$ nodes where there is a directed edge from node $m$ to node $n$ if $B_{nm}>0$. Then the matrix $B$ is irreducible if and only if its associated graph $G_{B}$ is strongly connected. In other words, $B$ is irreducible if there is a path between any two nodes of $G_{B}$. If $G_{B}$ is not strongly connected, we can separate it into strongly connected components and the threshold analysis which was presented in this section can be done on different components separately. Particularly, for an individual $i$ that never gets active we have $\gamma_{1}^{i}=0$ or equivalently $p_{2}^{i}=0$. In such a case we can see the node that corresponds to $I_{2}^{i}$ in the associated graph $G_{B}$ is disconnected from the rest of nodes and the threshold analysis can be carried out by eliminating the row and column for $I_{2}^{i}$ in the $J$ matrix. In fact, if in the matrix $J$ we exclude all those rows and columns that correspond to $I_{2}$ for the individuals that never gets active we can see the resulting matrix $B$ is irreducible if and only if union of the two layers, $\mathbb{L}_{1}$ and $\mathbb{L}_{2}$, is strongly connected.  

As we have shown, the threshold value $\beta^{\ast}$ is the smallest transmission rate $\beta$ for which the eigenvalue problem $J_{\beta}Z=0$ has a nontrivial solution. We can rewrite this eigenvalue problem as $B^{\star}Z=Z$, where
 \begin{figure*}[t]
	\subfloat[]{
		\includegraphics[width=.65\columnwidth]{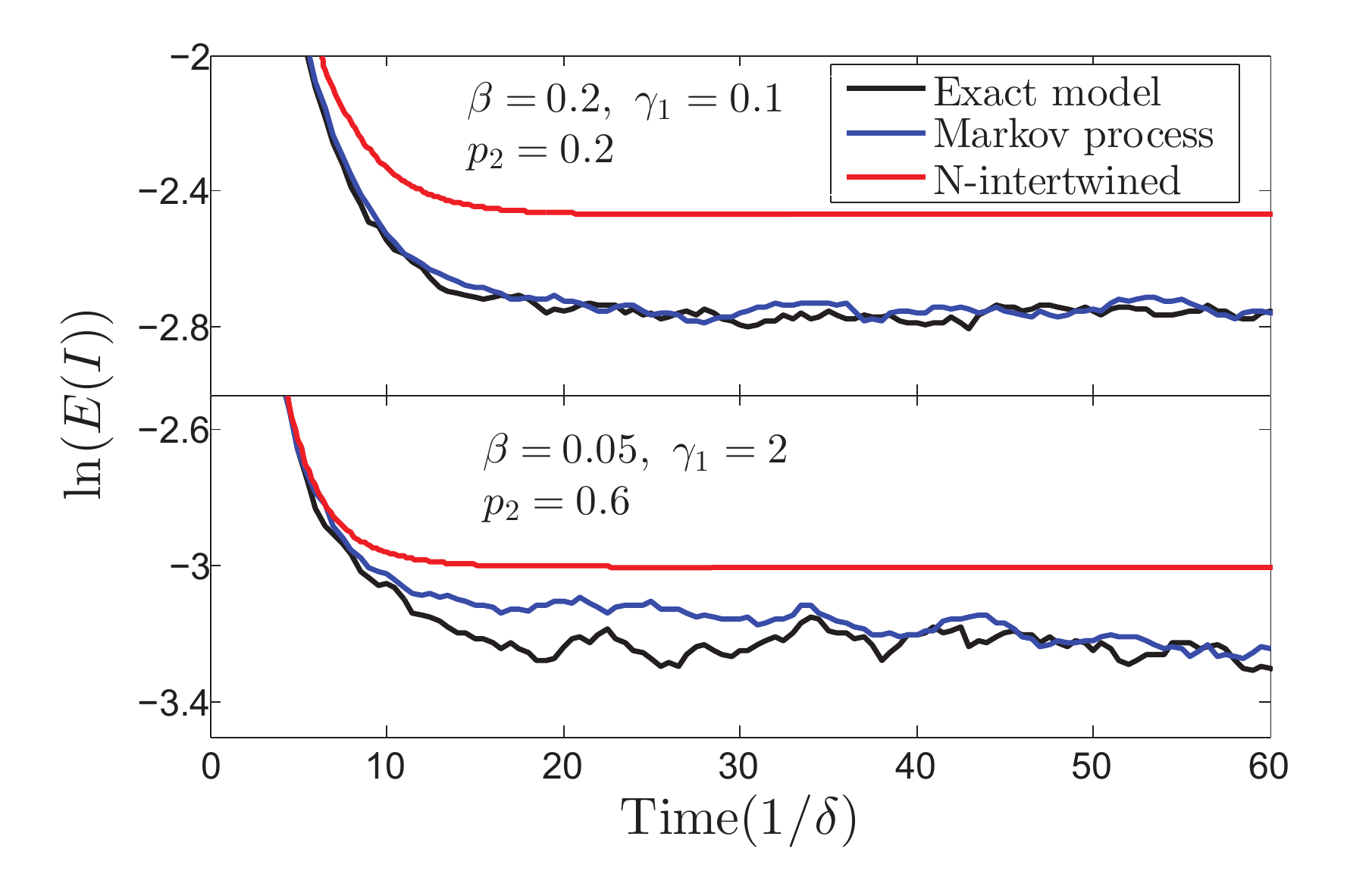}
\label{compsan}} 
	\subfloat[]{ 
	
		\includegraphics[width=.67\columnwidth]{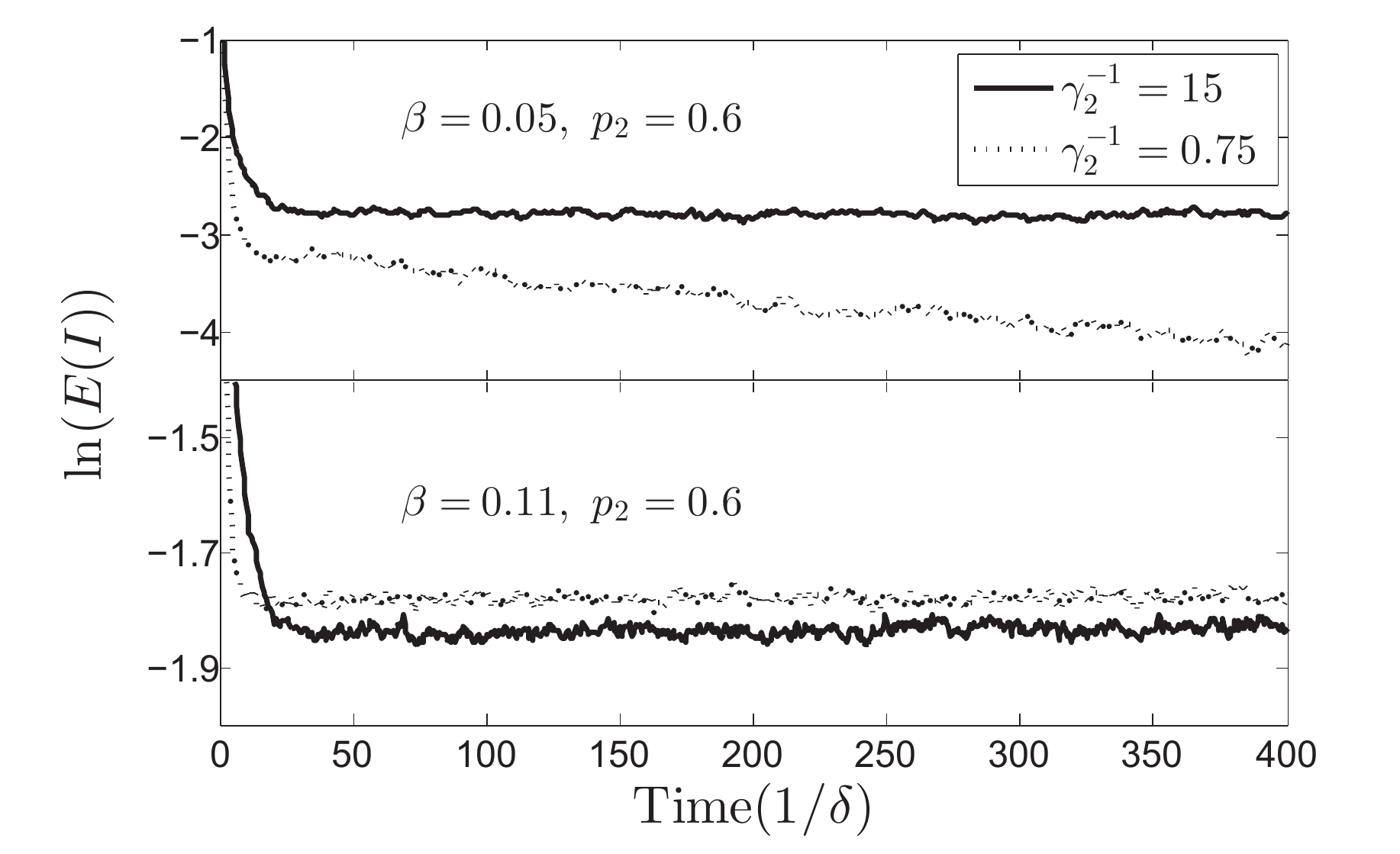}
\label{treshsan}}%
\subfloat[]{ 	
		\includegraphics[width=.69\columnwidth]{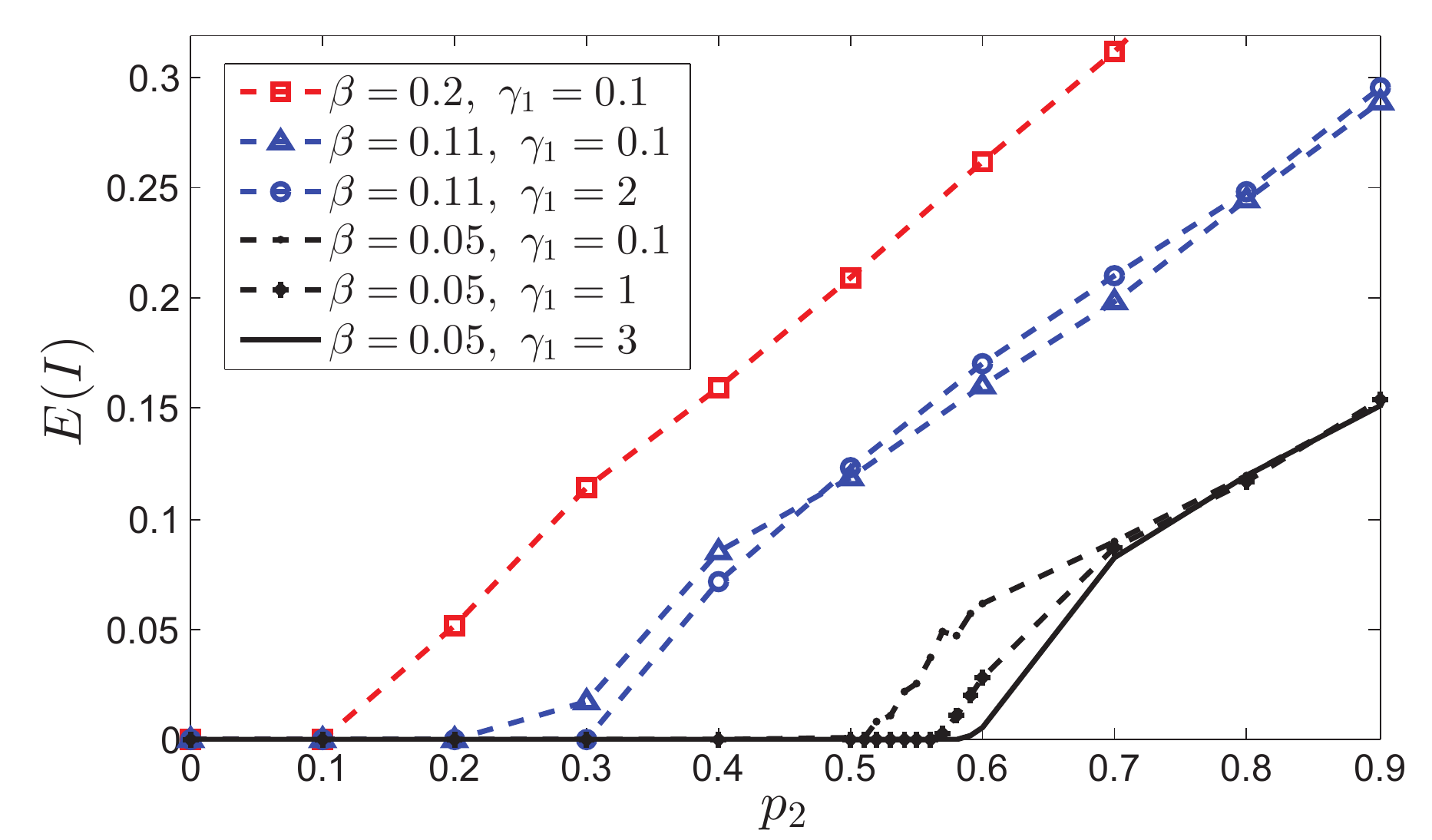}
\label{bifursan}}
	
	\caption{Results of the numerical and stochastic simulations discussed in section \ref{exp1}. Panel (a) compares the prevalence according to the two approximate processes and the exact spreading process; panel (b) shows the effect of link duration on the epidemic threshold in the exact process; panel (c) shows how the epidemic threshold is affected by different parameters in the exact process.} 
	\label{san}%
\end{figure*}
\begin{equation}\label{bstar}
\begin{split}
 B^{\star}=\begin{pmatrix}
 \frac{\beta}{\delta} A_{1} & \frac{\beta^{\prime}}{\delta} p_{2} A_{2}  \\
  \frac{\beta}{\delta} p_{2}^{\star}A_{1}+\gamma_{1}^{\star} & \frac{\beta^{\prime}}{\delta}p_{2}^{\star} A_{2}  
 \end{pmatrix},  
\end{split}
\end{equation}
and $\overline{p_{2}^{\star}}$, $\overline{\gamma_{1}^{\star}}$ are diagonal matrices such that
\[
(p_{2}^{\star})_{i,i}=\frac{\delta p_{2}^{i}}{\delta+\gamma_{1}^{i}+\gamma_{2}^{i}},\ \ \ (\gamma_{1}^{\star})_{i,i}=\frac{ \gamma_{1}^{i}}{\delta+\gamma_{1}^{i}+\gamma_{2}^{i}}.
\]

Since matrix $B^{\star}$ has the same structure as the matrix $B$, it is an irreducible matrix and its largest eigenvalue is positive. Hence, the threshold value $\beta^{\ast}$ is the transmission rate $\beta$ for which $\lambda_{\max}(B^{\star})=1$. 
\section{numerical results} 
In the following, we perform simulations to investigate the relation between the exact process and the mean-field approximation of the process. Moreover, we explore the effect of the model's parameters on the infection spreading.
\subsection{Experiments on a real-world network structure}\label{exp1}
In this section, we use the largest connected component of a network that represents sexual contacts among men who have sex with men in the city of San Francisco \cite{juher2017network}. This network has $953$ nodes and $1011$ links, where a few nodes with high degrees act as hubs. Although some of the links in this network are temporal, for this experiment, we treated all these links as the permanent contacts of the network layer $\mathbb{L}_{1}$. Next, since it is not possible to infer the potential contacts from the reported data in \cite{juher2017network}, we generated $\mathbb{L}_{2}$ as a synthetic network using the distance between the nodes in $\mathbb{L}_{1}$. It is possible to define different types of closeness for any two nodes in a connected graph like, for example, the shortest path distance or the resistance distance. Here we used the resistance distance and calculated the closeness of any two nodes in $\mathbb{L}_{1}$. To generate the neighborhood set in $\mathbb{L}_{2}$ of any node, $n$, we assumed that all the nodes with a distance to $n$ smaller than a threshold value are the neighbors of $n$, excluding those nodes that already have a permanent contact with $n$ in $\mathbb{L}_{1}$. Although, for this experiment we generated $\mathbb{L}_{2}$ using the closeness in the layer $\mathbb{L}_{1}$, in real-world applications we need to consider other types of relations between the nodes in the process of inferring potential contacts. One of these relations can be, for instance, the geographical distance. 

In the first experiment on this multilayer network, we compared the prevalence of infection obtained from three different processes discussed in the section \ref{comp}. We define prevalence as the average of nodal infection probabilities. As initial condition, we have assumed that all the nodes are active and infected at $t=0$. In figure \ref{compsan}, we have shown the prevalence as a function of time. In this figure, the curves referred to as ``N-intertwined'' show the prevalence calculated from the solution of equation \ref{eqa}. In the same figure, the curves that are labeled as ``Markov process'', are calculated using stochastic simulations. As we discussed in the section \ref{comp}, in this auxiliary process a potential contact in $\mathbb{L}_{2}$ transmits infection with probability $\beta p_0$ whenever the nodes at both ends of the link are active.  To estimate the prevalence at different points in time, we calculated the average of the infected population over $400$ simulations of the process. Finally, in figure \ref{compsan} we have also included the prevalence calculated using the stochastic simulations of the exact spreading model where the active nodes develop a contact over the potential links with probability $p_{0}=0.5$. The results of these simulations are the curves tagged as ``Exact model''. Based on our discussion in \ref{comp}, we expect that the prevalence obtained from the ``N-intertwined'' equations will be higher than that of the ``Markov process'' at anytime. Moreover, we also expect that the prevalence in the ``Markov process'' will be an upper-bound for the ``exact model''. We clearly see such a relation between the prevalence curves in figure \ref{compsan}. In fact, we repeated the simulations with different sets of parameters values and we observed the same trend.

In another experiment, we studied the effect of nodal activity rates on the prevalence of infection, when the exact spreading model is unfolding over the network. Figure \ref{treshsan} shows the curves obtained from the result of $400$ simulations. From this figure we can observe that, for $\beta=0.05$, the spreading process with $\gamma_{2}^{-1}=15$ reaches metastability, while in the process with $\gamma_{2}^{-1}=0.75$ infection dies out exponentially (note the logarithmic scale in the vertical axis). This might be counter-intuitive because, when the nodes changes the links too frequently, one may expect the infection spreads more easily. In contrast, in the simulations we observe that, for higher activity rates (with potential links becoming active and inactive more frequently), the threshold value is indeed higher. This can be explained by considering the infection process in the SIS model. For this model, we assume the infection transmission time is an exponential random variable with the expected value $\beta^{-1}$. When the link between a pair of nodes inactivates fast, the infection does not have enough time to be transmitted. In fact, in the simulation corresponding to figure \ref{treshsan}, after an initial period, all the nodes are active with the probability equal to $0.6$. Hence, the only difference between the curves with different activity rates is the duration of the links, which has an expected value of $(2\gamma_{2})^{-1}$. Although figure \ref{treshsan} shows that link duration can change the course of spreading process, more simulations reveal when nodes are active with high probabilities, the link duration is not as effective in the infection spreading as in the case where these nodal probabilities are low. When nodal probabilities of being active are high, if a node becomes active, then there is a high probability to encounter another active node and, hence, to develop a link that, in turn, increases the effective number of contacts.  In figure \ref{bifursan} we have plotted the (logarithm) infection prevalence in the meta-stable state as a function of $p_{2}$ (probability of being active). We can see that, for high values of the probability $p_{2}$, the prevalence curves for $\gamma_{1}=0.1$ and $\gamma_{1}=3$ with the same value of $\beta=0.05$ are almost similar to each other, while for lower values of $p_{2}$ they are different. In fact, since $p_{2}$ only depends on the ratio of $\gamma_{1}$ and $\gamma_{2}$, for a same value of $p_{2}$ the duration of links for $\gamma_{1}=0.1$ is 30 times higher than that of $\gamma_{1}=3$. In this figure, we can see that this difference between the values of the link duration is only significant for low values of $p_{2}$. 

\begin{figure}[t]	

		\includegraphics[width=1\columnwidth]{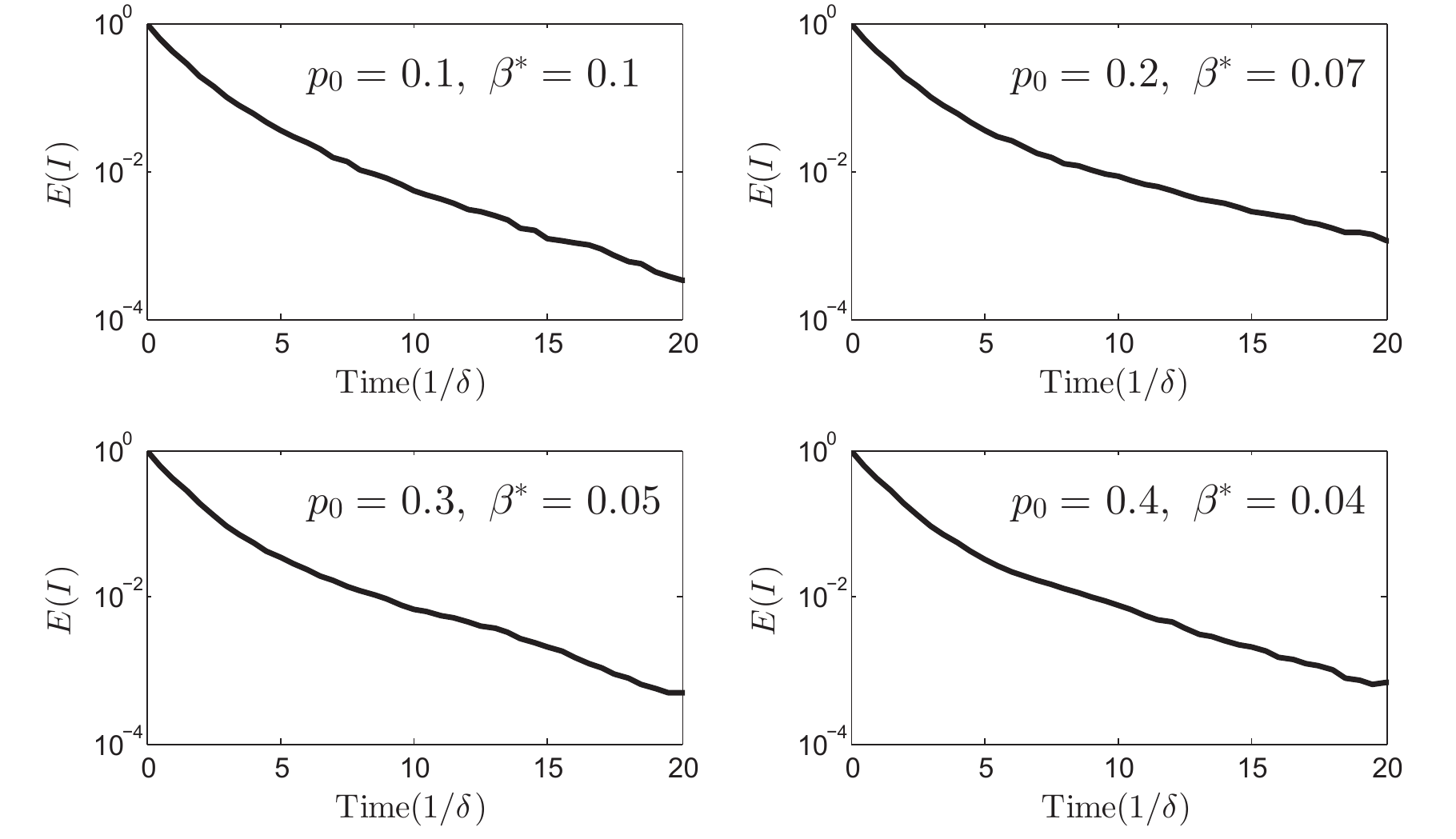}
	
	\caption{ This figure shows the approximate threshold $\beta^{\star}$ is a lower-bound for the exact epidemic threshold. Different plots show the prevalence of infection in the simulation of the exact process for different configurations in the parameters space. In all the simulations we have assumed $\beta$ is slightly lower than $\beta^{\star}$. Since, $\beta^{\star}$ provides a lower-bound for the exact epidemic threshold, we can see the infection dies out in all the simulations. } 
	\label{san2}%
\end{figure}

To investigate the relevance of the threshold $\beta^{\ast}$ obtained from the N-intertwined equations in section \ref{treshold}, we performed another set of simulations where for each node we exclusively assigned random values to $\gamma_{1}$ and $\gamma_{2}$. Next, we used the $B^{\star}$ matrix in equation \ref{bstar} to find the the threshold value $\beta^{\ast}$ for the transmission rate. Since the N-intertwined equations give an upper-bound for the nodal infection probabilities in the exact process, if the transmission rate $\beta$ is lower than $\beta^{\ast}$ we expect that the prevalence of infection in the exact process dies out. In figure \ref{san2}, we have plotted the result obtained from simulating the exact process for different configurations of parameters. In all the simulations, the transmission rate $\beta$ is slightly lower than the threshold value $\beta^{\ast}$ and we can see the infection is dying out. 

\begin{figure*}[t]
	\subfloat[]{
		\includegraphics[width=.67\columnwidth]{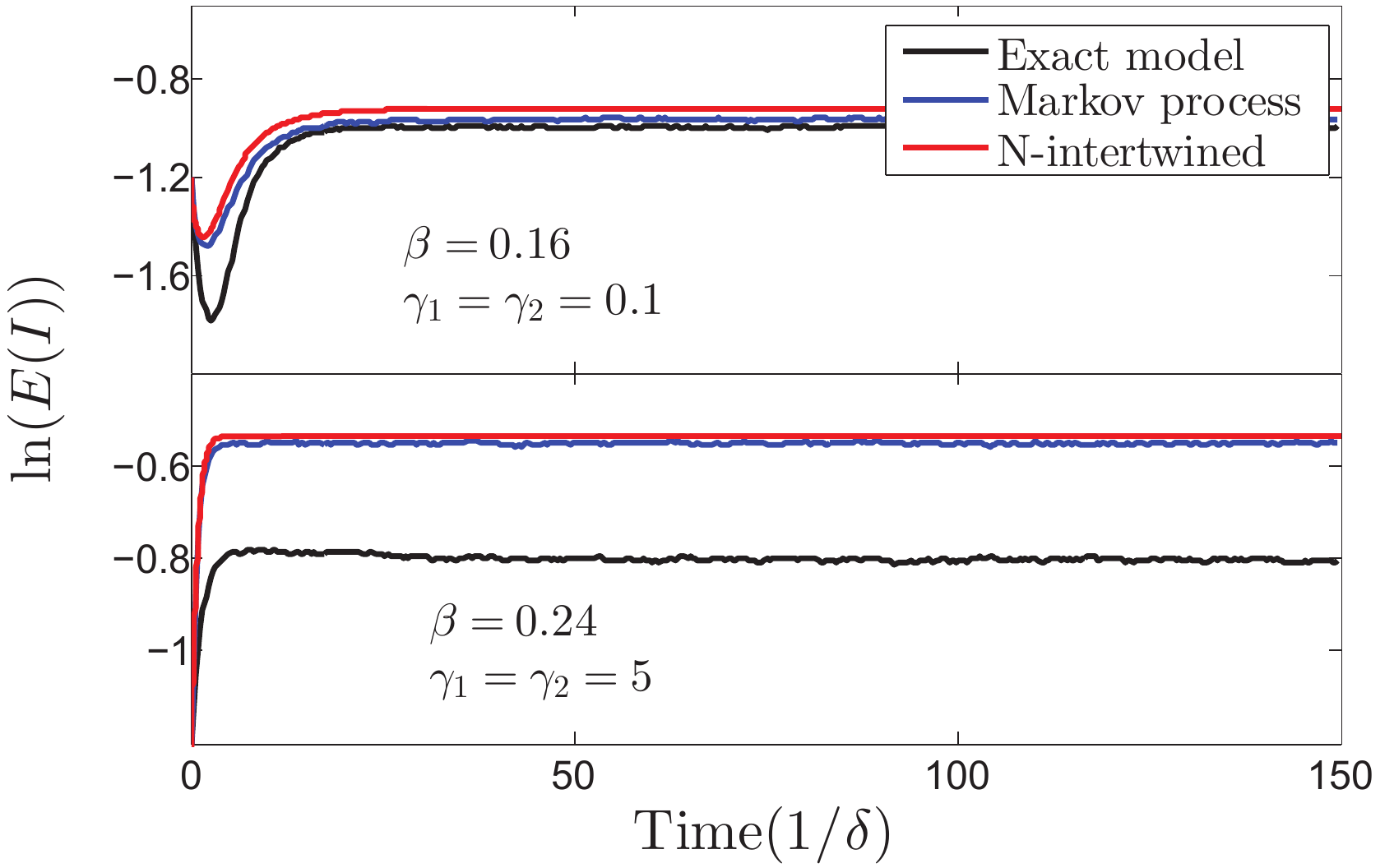}
\label{compRR}} 
	\subfloat[]{ 
	
		\includegraphics[width=.65\columnwidth]{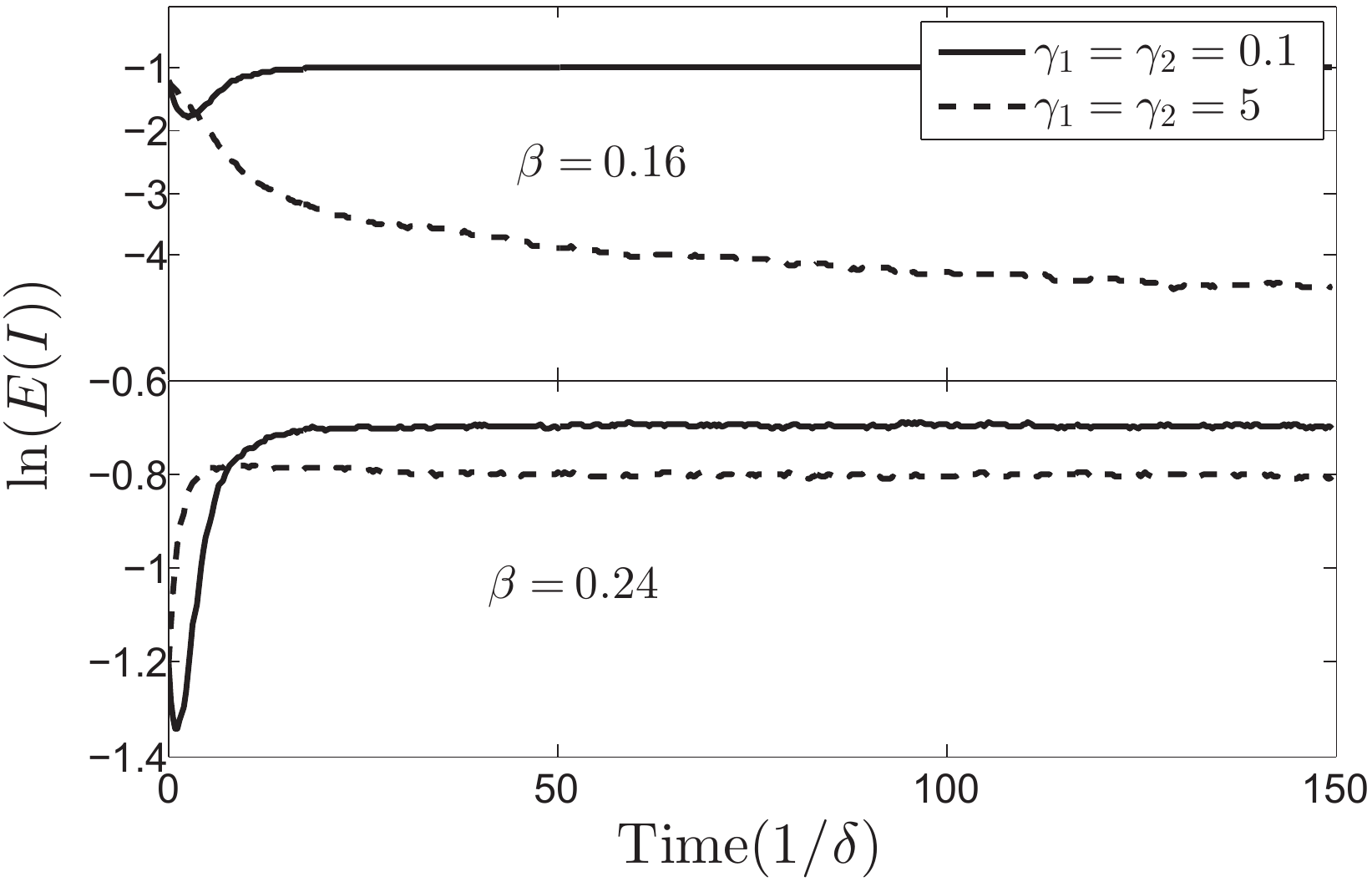}
\label{treshRR}}%
\subfloat[]{ 	
		\includegraphics[width=.68\columnwidth]{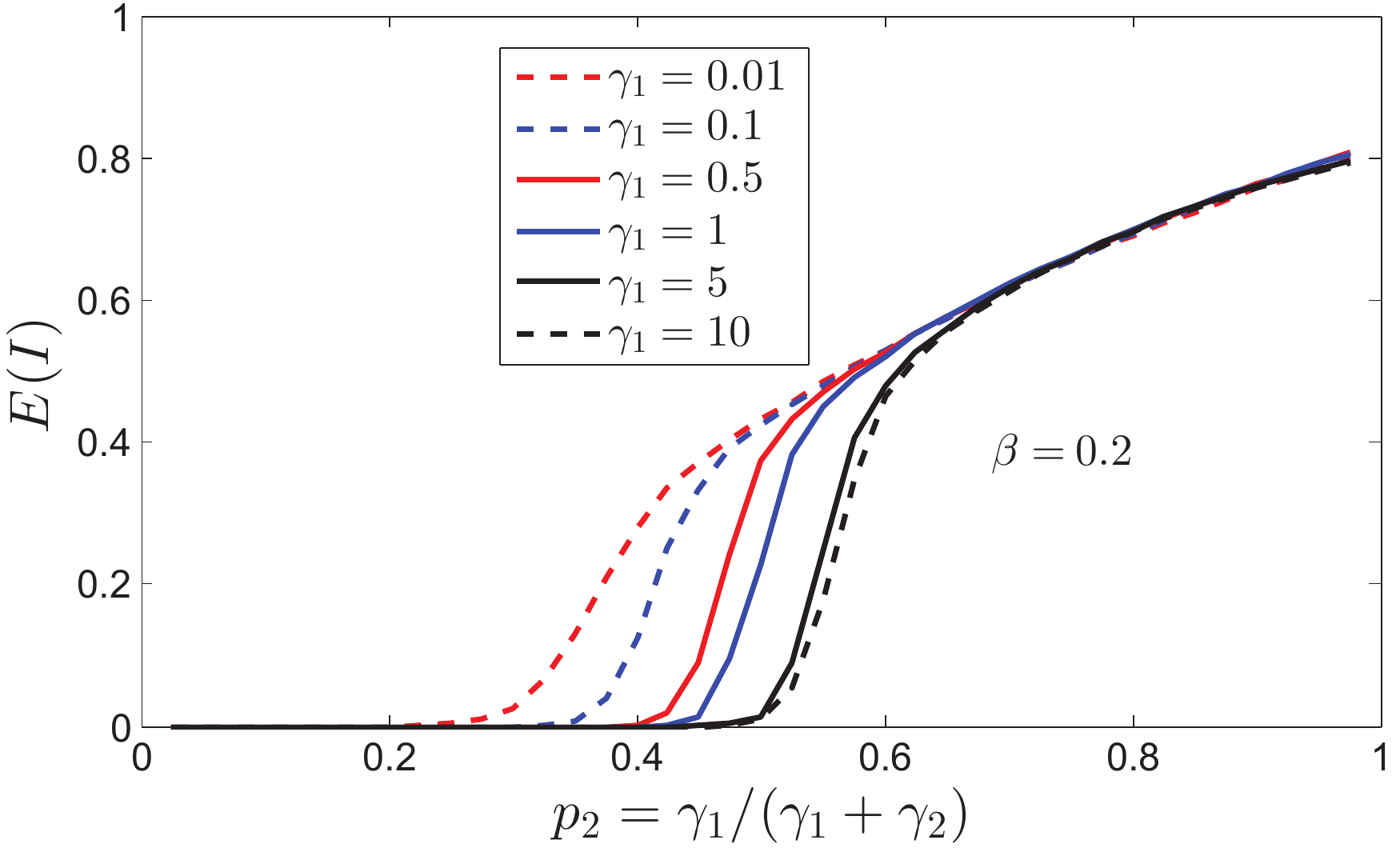}
\label{bifurRR}}
	
	 \caption{Results of numerical and stochastic simulations of the spreading processes on random regular graphs, discussed in section \ref{exp2}. Panel (a) shows the comparison of different approximate processes with the exact process; panel (b) shows the effect of link duration on the epidemic threshold in the exact process; panel (c) shows how the epidemic threshold is affected by different parameters in the exact process.}
	\label{RR}%
\end{figure*}
\subsection{Experiments on a random regular network}\label{exp2}
\begin{figure}[t]	
\centering
		\includegraphics[width=0.85\columnwidth]{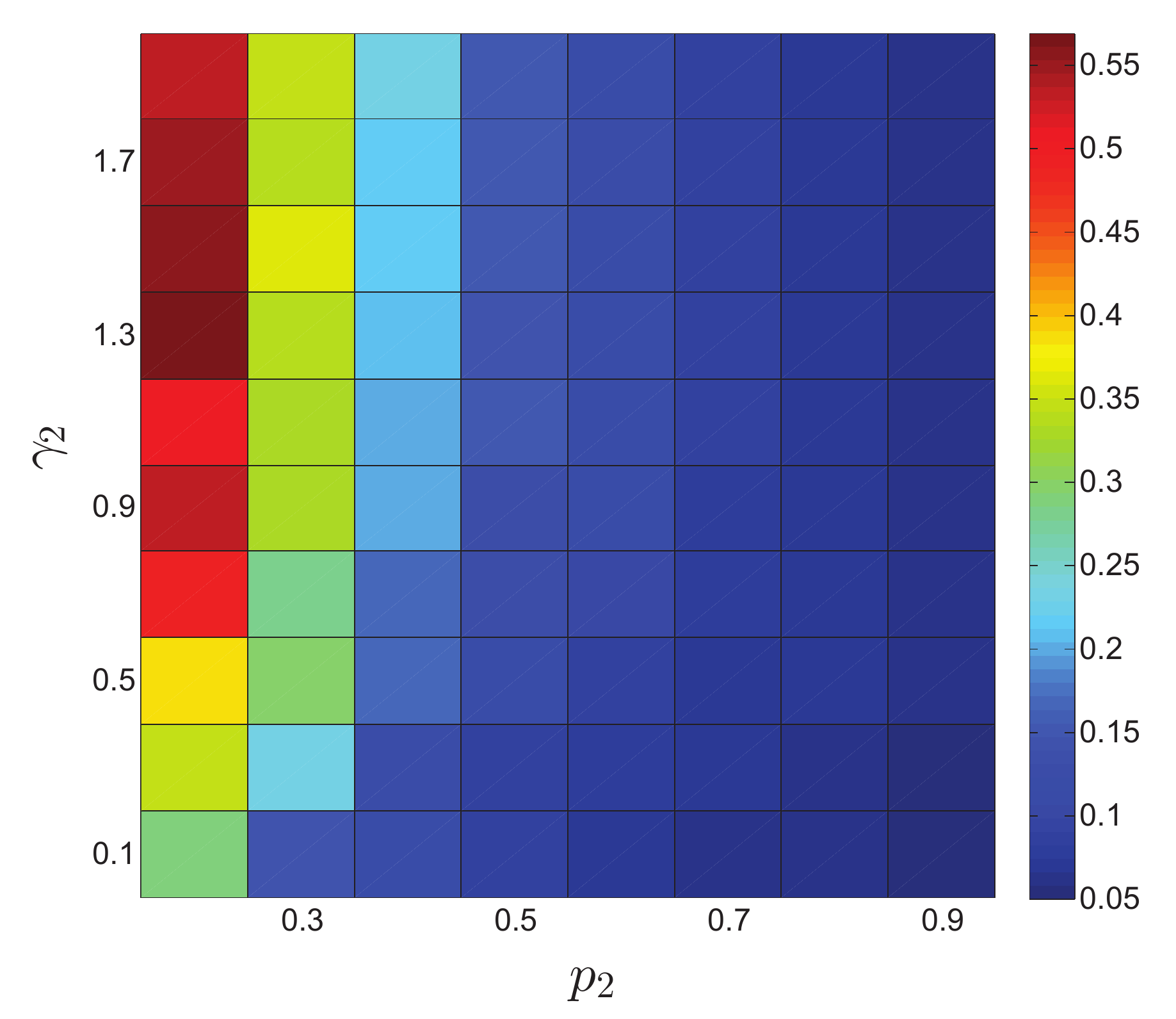}
	
	\caption{This figure shows the epidemic threshold, $\beta^{\star}$, obtained from the simulation of the exact spreading process. } 
	\label{heat}%
\end{figure}

To check the generality of numerical result in section \ref{exp1} we repeated the experiment on a random regular multilayer network of $500$ nodes. For the layer $\mathbb{L}_{1}$ we generated a random regular network where each node has four neighbors and for the layer $\mathbb{L}_{2}$ we used a  random regular network with the node degree of $50$. Figure \ref{compRR} shows the infection prevalence curves obtained from the N-intertwined approximation, the Markov process, and the exact spreading process. As we expect, the N-intertwined equations provide and upper-bound for the prevalence values obtained from the Markov process and the exact process. However, when the the nodal activity rates are high (bottom panel), the difference between the values of prevalence from the exact process and the N-intertwined equations is much higher. In figure \ref{treshRR}, we can observe the effect of the link duration on the epidemic threshold in the exact process. We can see that, when the mean duration of occasional links are low, $\gamma_{2}=5$, the metastable state happens at a higher value of $\beta$. This effect of the link duration in $\mathbb{L}_{2}$ on the epidemic threshold is also clear from figure \ref{bifurRR}. In this figure we see that reaching the metastable state requires less active nodes when the link duration is higher. However, given a value of $p_{2}$, if the transmission rate $\beta$ is far above the threshold (large prevalence), then the link duration does not significantly affect the prevalence. In figure \ref{heat}, we have shown the epidemic threshold, obtained from the simulation of the exact process, as a function of activity probability, $p_{2}$ and $\gamma_{2}$, which is proportional to the inverse of the link duration expectation. From this figure we can see when $p_{2}$ increases the threshold decreases. However, when the number of active nodes is small (lower value of $p_{2}$) the threshold increases when the duration of links decreases.
\section{conclusions}
In this work, we developed a novel temporal network model that incorporates nodal activities and pair-specific probabilities for developing links. Such model is suitable for studying infection spreading in real-world processes because it accounts for the possibility of establishing occasional contacts in addition to the permanent ones. Occasional contacts are modeled considering potential links that are activated with probability $p_0$ when both end-nodes are willing to develop new contacts. This model allows us to study the role of the potential contact layer on the spreading process by quantifying its utilization through the activity probability parameter $p_2$. In particular, we study how these different parameters can affect the metastable state of the infection spreading. By analyzing the SIS process over such network, we found a condition that guarantees the exponential die-out of infection. Moreover, we found that the prevalence of infection strongly depends on the utilization of the potential contact layer and the duration of occasional links, given a fixed value of the infection transmission rate. Our simulations show that, for a limited number of active nodes, the metastable state occurs when the duration of links increases. Conversely, for a high number of active nodes in the population, the duration of links is not very effective on the prevalence of infection. Overall, disregarding the potential contact layer can produce a non-negligible underestimation in the epidemic size prediction.
\ifCLASSOPTIONcompsoc
  \section*{Acknowledgments}
\else
  \section*{Acknowledgment}
\fi

This material is based on work supported by the National
Science Foundation under Grants No. DMS 1515810.

\ifCLASSOPTIONcaptionsoff
  \newpage
\fi

\bibliographystyle{IEEEtran}
\bibliography{Refrences}

\end{document}